\documentclass[twocolumn]{aastex631}
\usepackage{amsmath,amssymb,amsfonts,bm}
\usepackage{natbib}
\usepackage{graphicx}   
\usepackage{float}
\usepackage{longtable}
\usepackage{graphics}
\usepackage{hyperref}
\usepackage{color}
\usepackage{calc}
\usepackage{threeparttable}
\usepackage{ulem}

\newcommand \beq{\begin{equation}}
\newcommand \eeq{\end{equation}}
\newcommand \bey{\begin{eqnarray}}
\newcommand \eey{\end{eqnarray}}

\newcommand \kpc{\,{\rm kpc}}
\newcommand \mpc{\,{\rm Mpc}}

\newcommand \kms{\,{\rm km \, s}^{-1}}

\newcommand{\gsim}{\lower.5ex\hbox{$\; \buildrel > \over \sim \;$}}
\newcommand{\lsim}{\lower.5ex\hbox{$\; \buildrel < \over \sim \;$}}

\newcommand{\oii}{\hbox{[O\,{\sc ii}]}}
\newcommand{\oiii}{\hbox{[O\,{\sc iii}]}}

\def\Siiv{Si\,{\sc iv}}
\def\Civ{C\,{\sc iv}}
\def\Cii{C\,{\sc ii}}

\newcommand{\Oiiib}{[O\,{\sc iii}]\,$\lambda$5007}

\newcommand{\Ciii}{C\,{\sc iii]}}

\def\Mgii{Mg\,{\sc~ii}}
\def\Mgiiwave{Mg\,{\sc~ii}\,$\lambda$2798}
\newcommand{\Mgiia}{Mg{\sc ii}\,$\lambda$2796}
\newcommand{\Mgiib}{Mg{\sc ii}\,$\lambda$2803}
\newcommand{\Mgiiab}{Mg{\sc~ii}\,$\lambda\lambda$2796,2803}
\newcommand{\nvab}{N{\sc~V}\,$\lambda\lambda$1238,1242}

\newcommand{\lya}{Ly~$\alpha$}

\def\kms{$\rm km\,s^{-1}$}
\def\zabs{$z_{\rm abs}$}
\def\zem{$z_{\rm em}$}
\def\zemfg{$z_{\rm em}(fg)$}
\def\zembg{$z_{\rm em}(bg)$}
\def\Wr{$W_{\rm r}$}
\def\ergs{${\rm erg\,s^{-1}}$}

\shortauthors{Chen et al.}
\begin{document}
\title{The Study of Circumgalactic Medium with Quasar Pairs}
\author{Zhi-Fu Chen, Huan-Chang Qin, Jin-Ting Cai, Yu-Tao Zhou, Zhe-Geng Chen, Ting-Ting Pang, Zhi-Wen Wang}

\email{zhichenfu@126.com, hcqin@bsuc.edu.cn}

\footnotetext{School of Mathematics and Physics, Guangxi Minzu University, Nanning 530006, China}
%
%
%
%
%
%
%

\begin{abstract}
We have collected 10025 foreground-background quasar pairs with projected distances $d_p<500$ \kpc\ from the large quasar catalog of the SDSS DR16Q. We investigate the properties of the \Mgii\ absorption lines with $W_r>0.15$ \AA\ around foreground quasars, including both the LOS (line-of-sights of foreground quasars) and transverse (TRA, perpendicular to the LOS) absorptions. Both the equivalent width (the correlation coefficient $\rho=-0.915$ and the probability $P < 10^{-4}$ of no correlation) and incident rate ($\rho=-0.964$ and $P < 10^{-6}$) of TRA \Mgii\ absorption lines are obviously anti-correlated with projected distance. The incident rate of TRA \Mgii\ absorption lines is obviously ($>4\sigma$) greater than that of LOS \Mgii\ absorption lines at projected distances $d_p<200$ \kpc, while the TRA and LOS \Mgii\ both have similar ($<3\sigma$) incident rates at scales $d_p>200$ \kpc. The anisotropic radiation from quasars would be the most possible interpretation for the anisotropic absorption around quasars. This could also indicate that the quasar radiation is not obviously impacting the gas halos of quasars at scales $d_p>200$ \kpc.
\end{abstract}

\section{Introduction}
Since absorption lines were observed on quasar spectra, a great effort has been made to understand the link between absorbing gas and foreground galaxies approaching to the quasar sightline \cite[e.g.,][]{1966ApJ...144..847B,1969ApJ...156L..63B,1974Natur.250..310O,1975A&A....45..329R,1977ApJ...218...33B,1978PhyS...17..229B}. The absorbing gas content is likely related to the global properties of galaxies. The number density of absorbers and star formation rate density of galaxies show similar evolution profile \cite[e.g.,][]{2006ApJ...639..766P,2012ApJ...761..112M,2013ApJ...763...37C,2013RAA....13..641C,2013ApJ...770..130Z,2014ARA&A..52..415M,2015ApJS..221...32C}, implying that the star formation rate within galaxy would be an important factor giving rise to absorption line. A more positive evidence for the relationship between absorption line and star formation rate is that the \Mgii\ with larger equivalent widths tend to host \oii\ with stronger emission flux. \cite[e.g.,][]{2011MNRAS.417..801M,2011MNRAS.412.1559N,2012ApJ...748..131S,2018MNRAS.476..210J}. The luminosity of a galaxy plays a role in the spatial distribution of gas in its dark matter halo \cite[][]{2008ApJ...679.1218T}, and the strength of absorption line is dominated by the amount of gas, therefore the properties of absorption lines are found to be related to the luminosity and stellar mass of galaxy \cite[e.g.,][]{2008ApJ...687..745C,2010ApJ...714.1521C,2010ApJ...724L.176C,2011ApJ...740...30L,2013ApJ...763L..42C,2021ApJ...923...56D,2021MNRAS.502.4743H}. Morphology, inclination, azimuthal angle with respect to the major axis of a galaxy, and environment of galaxies also influence the quantity of gas that the quasar sightline intercepts, therefore the connection between them and the properties of absorption lines are widely investigated\cite[e.g.,][]{2007ApJ...662..909K,2011MNRAS.416.3118K,2011ApJ...743...10B,2014MNRAS.438.2530C,2014PASP..126..969B,2017ARA&A..55..389T,2018ApJ...866...36L,2020ApJ...904...76H,2021MNRAS.503.4309L}. In the last years, blind searches for galaxies associated with absorbers also revealed that the absorber's properties are related to the gravitational interactions in the group populations, galaxy's outflows, intergalactic medium accretions, star formation rates, and gas distributions within galaxy group\cite[e.g.,][]{2019MNRAS.490.1451F,2020MNRAS.493.5336B,2020MNRAS.499.5022D,2020ApJ...904..164N,2021MNRAS.508.5612M,2022MNRAS.514.6074N,2022arXiv220915021L}. In addition, one absorber might be associated with multiple galaxies within a group as well\cite[e.g.,][]{2020MNRAS.493.5336B,2022arXiv220915021L}.

The activity in nuclei is an important factor regulating the distribution and amount of galaxy gas. The integral field units (IFU) is a good method to investigate the kinematic properties across the entire galaxies \cite[e.g.,][]{2014A&A...568A..70B,2015AJ....150...19L,2021ApJ...923....6J,2022MNRAS.tmp.1897Z}. Making full use of the absorption lines originated in gaseous halo of normal galaxies and imprinted in the spectra of background objects, considerable studies\cite[e.g.,][]{2016MNRAS.455.1713H,2018ApJ...866...36L,2020ApJ...904...76H,2021MNRAS.504...65A,2021MNRAS.502.4743H} revealed that the absorption line properties related to the structure of gaseous halo, star formation activity, morphology, environment of the galaxy. Meanwhile over the last 20 years, noticeable attentions\cite[e.g.,][]{2006ApJ...645L.105B,2006ApJ...651...61H,2014MNRAS.441..886F,2015MNRAS.452.2553J} are also paid to study the gaseous halo of the galaxy existing a quasar in its center with its absorptions to the background quasars, and uncovered that the gas extension, distribution, and content are related to the activities within the galaxy centre. It is widely accepted that the circumgalactic medium (CGM) surrounding quasar is a reservoir of enrich gas \cite[e.g.,][]{2013ApJ...762L..19P,2015MNRAS.452.2553J}. The study of quasar CGMs is a good chance to investigate the surrounding environment and feedback of quasar. In the last decade, many works have also been focused on the emissions from CGMs \cite[e.g.,][]{2013ApJ...766...58H,2014Natur.506...63C,2014ApJ...786..106M,2015Natur.524..192M,2015Sci...348..779H,2016ApJ...831...39B,2017ApJ...848...78F,2018MNRAS.476.2421G,2018MNRAS.473.3907A,2019MNRAS.482.3162A,2019ApJS..245...23C,2019MNRAS.483.5188C,2019ApJ...881..131D,2019ApJ...887..196F,2020ApJ...894....3O,2020ApJ...898...26G,2021ApJ...909..151B,2021MNRAS.503.3044F}.
The investigations in emissions are redshift dependent and might spend a lot of exposure time of large-aperture telescopes, but a single exposure can image the entirety of the nebular emission. Although the detection of absorption features is limited to a single (or a few at max) location in the gas halo of a galaxy, it is redshift independent. Thus, the detections of emissions and absorptions from gas halo of the galaxy are the complementary approaches.

Absorption lines with redshifts close to quasar systemic redshifts (\zabs$\approx$\zem, associated absorption line) are likely formed in gaseous cloud physically associated with quasar themselves. Large samples\citep[e.g.,][]{2012ApJ...748..131S,2016MNRAS.462.2980C,2017ApJS..229...22H,2018ApJS..235...11C,2018ApJS..239...23C,2020ApJS..250....3C,2022A&A...659A.103C} of associated absorption lines have been available through systematically analyzing the quasar spectra from the Sloan Digital Sky Survey \citep[SDSS,][]{2000AJ....120.1579Y}, which can often be used to comprehend the gas content of quasar systems along sightlines. The major shortcoming of associated absorption lines along quasar sightlines is that they cannot provide any information of gas content located on the transverse direction of quasars. If the sightline of a background luminous quasar passes through the vicinal medium of a foreground quasar, one can make use of the absorption features imprinted on the background quasar spectra to limit more comprehensively the gas content and environment surrounding the foreground quasar.

The supermassive black hole and powerful outflows/jets of quasars might significantly affect surrounding environments via strong feedbacks \citep[e.g.,][]{2022NatAs...6..339C,2022MNRAS.513.1141Z}, leading to the properties of quasar CGMs substantially differ from those of normal galaxies. The strong radiation originated from quasar might photoionize, and even photoevaporate surrounding gas from tens \kpc\ to a few \mpc\
\cite[e.g.,][]{2008MNRAS.388..227W,2009MNRAS.392.1539T,2014ApJ...796..140P,2015MNRAS.452.2553J,2019ApJ...884..151J,2022SciA....8.3291H}. In addition, the strong directional jet probably gives rise to anisotropic gas halo around quasar \cite[e.g.,][]{1993ARA&A..31..473A,2000ApJ...545...63E,2007ApJ...655..735H,2014MNRAS.441..886F,2017ApJ...843....3A,2019ApJ...884..151J}. Along the sightlines of quasars, the powerful outflows/jets likely over ionize the gas medium with low ionization states and close to quasar central regions. While along the transverse direction, the gas medium is likely much less illuminated by the radiation from quasars and maintained low ionization state. These would be the most possible explanation to the proximity effect along the quasar's sightline and inverse proximity effect along transverse directions of quasars \cite[e.g.,][]{2007ApJ...655..735H,2019ApJ...884..151J}\footnote{In the vicinity of quasars, one expects that the amount of absorption lines should be increased with decreasing distance from quasar central regions. However, the strong radiation from quasars reduces the detectable absorption lines with lower ionization state as the absorption redshifts closer to the emission redshifts of the quasars. This phenomenon is commonly called as proximity effect \citep[e.g.,][]{1982MNRAS.198...91C,1996MNRAS.280..767S,2013ApJ...776..136P}.}. Here, the inverse proximity effect indicates that the number of absorption lines is increased with decreasing distance from quasar central regions.

Recently, as more and more quasars are discovered, the technique that quasars probe quasars is widely used to investigate the surrounding gas of quasars with absorption lines \cite[e.g.,][]{2006ApJ...645L.105B,2006ApJ...651...61H,2007ApJ...655..735H,2009MNRAS.392.1539T,2013MNRAS.429.1267F,2013ApJ...766...58H,2013ApJ...776..136P,2014MNRAS.441..886F,2014ApJ...796..140P,2014MNRAS.444.1835S,2015MNRAS.452.2553J,2015MNRAS.446.2861K,2016MNRAS.457..267L}.
It is almost without exception that all the neutral hydrogen \lya, and metal \Cii, \Civ\ and \Mgii\ absorptions exhibit an excessive incidence over cosmic average on transverse direction, and the incident rate of absorptions on transverse direction is inversely  proportional to the projected distance from foreground quasars. In this paper, making use of the final quasar catalog from the SDSS, we construct a large sample of foreground-background quasar pairs to statistically analyze the properties of \Mgiiab\ absorbing gas around foreground quasars, including the absorptions in the line-of-sight (LOS absorptions) and transverse (TRA absorptions) directions of foreground quasars. Making use of the LOS and TRA \Mgii\ absorption lines, ones can examine more thoroughly the gas environment surrounding quasars.

Throughout this work, we assume a flat $\Lambda$CDM cosmology with $\Omega_m = 0.3$, $\Omega_\Lambda = 0.7$, and $h_0 = 0.7$.\

\section{Sample selection}\label{sect:qsoselection}
The SDSS is an ambitious project to image the sky using a dedicated 2.5 m wide-field telescope \cite[][]{2006AJ....131.2332G} located at Apache Point Observatory in New Mexico, United States. During the first and second stages \citep[SDSS-I/II, from 2000 --- 2008; ]{2009ApJS..182..543A}, the SDSS obtained quasar spectra in a coverage wavelength from 3800 \AA\ to 9200 \AA\ with $R\equiv\lambda/\Delta\lambda\approx2000$. During the third and fourth stages \citep[SDSS-III/IV, from 2008 --- 2020; ][]{2013AJ....146...32S,2013AJ....145...10D,2016AJ....151...44D}, the SDSS gathered quasar spectra in a coverage wavelength from 3600 \AA\ to 10500 \AA\ at a resolution of $R\approx1300\sim2500$. The Sixteenth Data Release is the final dataset for the SDSS-IV quasar catalog\citep[DR16Q;][]{2020ApJS..250....8L}, which includes 750,414 quasars that are accumulated from SDSS-I to SDSS-IV. Using the quasar spectra from DR16Q, we construct a large sample of quasar pairs to investigate the properties of \Mgiiab\ absorption lines around foreground quasars. We select quasars from DR16Q with following criteria.
\begin{enumerate}
  \item The projected distance of quasar pairs at foreground quasar redshifts are limited to $d_p<500$ \kpc. Note that the 500 \kpc\ is a conservative value. The halo size of a massive galaxy might be less than 200 --- 300 \kpc, and the gas located at a distance $>500$ \kpc\ from the quasar center region is likely related to the halo of satellite galaxy \citep[e.g.,][]{2014MNRAS.439.3139Z}
  \item Narrow absorption lines are often confused by \lya\ forest absorptions if they fall into the spectral region with $(1+z_{\rm em})\times\lambda_{\rm Ly \alpha}>$$(1+z_{\rm abs})\times\lambda_{abs}$, where the $\lambda_{\rm Ly \alpha}$ and $\lambda_{\rm abs}$ are rest-frame wavelengths of the \lya\ emission line and detected absorption line, respectively. The absorptions of \nvab\ are often presented in the red wings of \lya\ emission lines, which would lead to misidentifications \Mgiiab\ absorption doublets. Therefore, we conservatively cut off parent sample to quasars with $(1+z_{\rm em}(fg))\times2800$\AA\ $\ge(1+z_{\rm em}(bg))\times1300$\AA, where \zemfg\ and \zembg\ are systematical redshifts of foreground and background quasars, respectively.
  \item Accounting for the rest-frame wavelength of \Mgii\ absorption lines and the coverage wavelengths of the SDSS spectra, we only consider the quasars with $z_{em}>0.4$. In addition, in order to avoid the contamination from outflows of background quasars, the differences between the redshifts of foreground and background quasars $\Delta\upsilon(z_{em}(fg),z_{em}(bg))>6000$ \kms.
  \item The LOS \Mgii\ absorption lines are detected in the spectral data around the \Mgii\ emission lines of foreground quasars. Thus, we only select the foreground quasars with median signal-to-noise ratio $\rm S/N >3$ within the spectral regions of $\rm \pm 6000$ \kms\ around \Mgii\ emission lines.
  \item The TRA \Mgii\ absorption lines are detected in the background quasar spectra. Thus, we also limit the background quasars with median signal-to-noise ratio $\rm S/N >3$ within the spectral regions of $\rm  \pm 6000$ \kms\ around foreground quasar redshifts.
\end{enumerate}
Above limits yield a sample of 10025 foreground-background quasar pairs. Using bolometric luminosity correction factor \cite[e.g.,][]{2011ApJS..194...45S}, we also compute the bolometric luminosity of the quasar from the flux density of the pseudo-continuum fitting at 3000 \AA, or 1350 \AA\ when the spectral data at 3000 \AA\ are not available for high redshift quasars. The parameters of quasar pairs are listed in Table \ref{tab:fgabs} and \ref{tab:bgabs}.

\begin{table*}
\caption{The properties of the quasar pairs and the LOS \Mgii\ absorption lines} \tabcolsep 0.3mm \small
\centering
\label{tab:fgabs}
 \begin{tabular}{ccccccccccccccccc}
 \hline\hline\noalign{\smallskip}
Pair ID&  SDSS name & PlateID & MJD & FiberID & $z_{\rm em}$& $\rm Log L_{bol}$& $d_{\rm p}$&$z_{\rm abs}$ & $W_{\rm r}^{\lambda2796}$ & $\sigma_{W_{\rm r}^{\lambda2796}}$& $W_{\rm r}^{\lambda2803}$&$\sigma_{W_{\rm r}^{\lambda2803}}$&  SDSS name \\
&foreground & & & &&\ergs& \kpc & &\AA&\AA&\AA&\AA&background \\
\hline\noalign{\smallskip}
1	&	154530.23+484608.9	&	8429	&	57893	&	896	&	0.4009 	&	46.304	&	459.0 	&	-	&	-	&	0.02	&	-	&	0.02	&	154535.84+484713.6	\\
2	&	081107.57+342940.8	&	9355	&	57814	&	84	&	0.4000 	&	44.609	&	422.1 	&	-	&	-	&	0.15	&	-	&	0.15	&	081104.05+342835.5	\\
3	&	075051.72+245409.3	&	928	    &	52578	&	232	&	0.4003 	&	46.174	&	475.0 	&	-	&	-	&	0.04	&	-	&	0.04	&	075046.78+245311.9	\\
4	&	114436.33+464232.9	&	1444	&	53054	&	103	&	0.4005 	&	45.399	&	401.6 	&	-	&	-	&	0.12	&	-	&	0.12	&	114429.71+464202.6	\\
5	&	104859.47+313326.1	&	10472	&	58159	&	936	&	0.4016 	&	44.944	&	274.8 	&	-	&	-	&	0.09	&	-	&	0.09	&	104903.42+313318.3	\\
6	&	094059.64+491618.3	&	7292	&	56709	&	446	&	0.4014 	&	44.518	&	440.8 	&	-	&	-	&	0.25	&	-	&	0.25	&	094105.87+491523.7	\\
7	&	081931.20+334837.5	&	9358	&	57749	&	212	&	0.4021 	&	46.243	&	484.4 	&	-	&	-	&	0.02	&	-	&	0.02	&	081928.29+334959.7	\\
8	&	083026.01+534834.0	&	7277	&	56748	&	350	&	0.4030 	&	44.501	&	450.7 	&	-	&	-	&	0.26	&	-	&	0.26	&	083021.29+534721.8	\\
9	&	221132.77+241955.4	&	7643	&	57302	&	356	&	0.4031 	&	44.751	&	494.4 	&	-	&	-	&	0.16	&	-	&	0.16	&	221127.39+242049.9	\\
10	&	234252.37-003915.3	&	9209	&	57686	&	655	&	0.4049 	&	45.172	&	481.6 	&	-	&	-	&	0.06	&	-	&	0.06	&	234256.67-003814.0	\\
11	&	013457.91+002738.2	&	4230	&	55483	&	804	&	0.4048 	&	44.394	&	422.0 	&0.4054	&	0.98&	0.17	&	1.25&	0.41	&	013502.28+002656.0	\\
\noalign{\smallskip}
\hline\hline\noalign{\smallskip}
\end{tabular}
\begin{flushleft}
Note --- Symbols ``-'' represent the spectra without detected \Mgii\ absorption lines. Only the equivalent width of $1\sigma$ upper limit is provided for the spectra without detected \Mgii\ absorption lines.
\end{flushleft}
\end{table*}

\begin{table*}
\caption{The background quasars with TRA \Mgii\ absorption lines} \tabcolsep 1.3mm \small
\centering
\label{tab:bgabs}
 \begin{tabular}{ccccccccccccccccc}
 \hline\hline\noalign{\smallskip}
Pair ID&  SDSS name & PlateID & MJD & FiberID & $z_{\rm em}$& $d_{\rm p}$&$z_{\rm abs}$ & $W_{\rm r}^{\lambda2796}$ & $\sigma_{W_{\rm r}^{\lambda2796}}$& $W_{\rm r}^{\lambda2803}$&$\sigma_{W_{\rm r}^{\lambda2803}}$ \\
 & & & & && \kpc & &\AA&\AA&\AA&\AA \\
\hline\noalign{\smallskip}
1	&	154535.84+484713.6	&	812	&	52352	&	360	&	1.4093 	&	459.0 	&	-	&	-	&	0.27 	&	-	&	0.27 	\\
2	&	081104.05+342835.5	&	9355	&	57814	&	88	&	1.7466 	&	422.1 	&	-	&	-	&	0.27 	&	-	&	0.27 	\\
3	&	075046.78+245311.9	&	11104	&	58436	&	142	&	1.1430 	&	475.0 	&	-	&	-	&	0.26 	&	-	&	0.26 	\\
4	&	114429.71+464202.6	&	7401	&	56808	&	654	&	1.5080 	&	401.6 	&	-	&	-	&	0.30 	&	-	&	0.30 	\\
5	&	104903.42+313318.3	&	6445	&	56366	&	604	&	0.8480 	&	274.8 	&	-	&	-	&	0.09 	&	-	&	0.09 	\\
6	&	094105.87+491523.7	&	7292	&	56709	&	450	&	1.7984 	&	440.8 	&	-	&	-	&	0.26 	&	-	&	0.26 	\\
7	&	081928.29+334959.7	&	3759	&	55236	&	190	&	0.7810 	&	484.4 	&	-	&	-	&	0.13 	&	-	&	0.13 	\\
8	&	083021.29+534721.8	&	7277	&	56748	&	346	&	1.5672 	&	450.7 	&	-	&	-	&	0.17 	&	-	&	0.17 	\\
9	&	221127.39+242049.9	&	7643	&	57302	&	354	&	1.1602 	&	494.4 	&	-	&	-	&	0.16 	&	-	&	0.16 	\\
10	&	234256.67-003814.0	&	9209	&	57686	&	659	&	1.8035 	&	481.6 	&	-	&	-	&	0.26 	&	-	&	0.26 	\\
11	&	013502.28+002656.0	&	697	&	52226	&	621	&	1.4232 	&	422.0 	&	-	&	-	&	0.17 	&	-	&	0.17 	\\
12	&	131144.26+420411.3	&	6620	&	56368	&	312	&	0.6550 	&	450.2 	&	-	&	-	&	0.11 	&	-	&	0.11 	\\
13	&	145310.47+420034.8	&	1397	&	53119	&	576	&	0.8046 	&	261.3 	&	-	&	-	&	0.17 	&	-	&	0.17 	\\
14	&	151928.77+573807.2	&	612	&	52079	&	548	&	1.4930 	&	420.0 	&	-	&	-	&	0.12 	&	-	&	0.12 	\\
15	&	091517.07+332155.7	&	10242	&	58161	&	573	&	1.8943 	&	363.4 	&	-	&	-	&	0.16 	&	-	&	0.16 	\\
16	&	075204.01+345711.3	&	9354	&	57806	&	296	&	1.7245 	&	451.8 	&	-	&	-	&	0.12 	&	-	&	0.12 	\\
17	&	100605.21+351159.9	&	4637	&	55616	&	130	&	1.4750 	&	352.8 	&	-	&	-	&	0.22 	&	-	&	0.22 	\\
18	&	025257.22-072217.4	&	457	&	51901	&	429	&	1.4488 	&	344.7 	&	-	&	-	&	0.21 	&	-	&	0.21 	\\
19	&	090558.02+411725.7	&	4605	&	55971	&	830	&	0.6038 	&	414.1 	&	-	&	-	&	0.16 	&	-	&	0.16 	\\
20	&	085850.21+514655.1	&	5155	&	55946	&	174	&	2.0286 	&	303.0 	&	-	&	-	&	0.15 	&	-	&	0.15 	\\
21	&	014729.24+002000.7	&	7843	&	56902	&	934	&	1.5188 	&	254.1 	&	0.4053	&	1.75	&	0.50 	&	0.65	&	0.24 	\\
\noalign{\smallskip}
\hline\hline\noalign{\smallskip}
\end{tabular}
\begin{flushleft}
Note --- Symbols ``-'' represent the spectra without detected \Mgii\ absorption lines. Only the equivalent width of $1\sigma$ upper limit is provided for the spectra without detected \Mgii\ absorption lines.
\end{flushleft}
\end{table*}

\subsection{Quasar redshifts}
\label{sect:redshift}
\begin{figure}[!ht]
  \centering
  \includegraphics[width=0.47\textwidth]{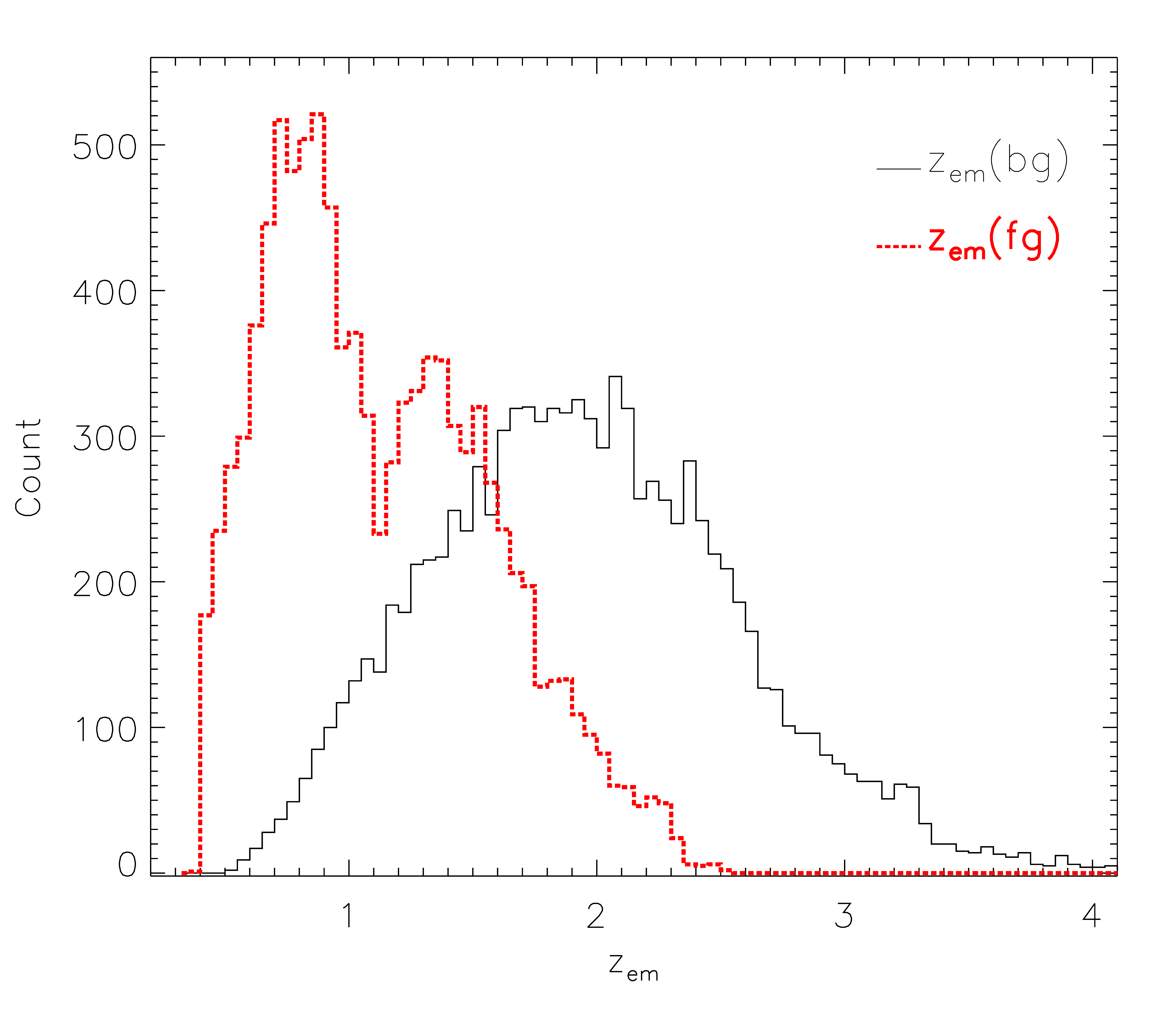}
  \caption{Redshift distributions. Black solid-line and red dash-line are for background and foreground quasars, respectively.}
  \label{fig:zem}
\end{figure}

\begin{figure}[!ht]
  \centering
  \includegraphics[width=0.47\textwidth]{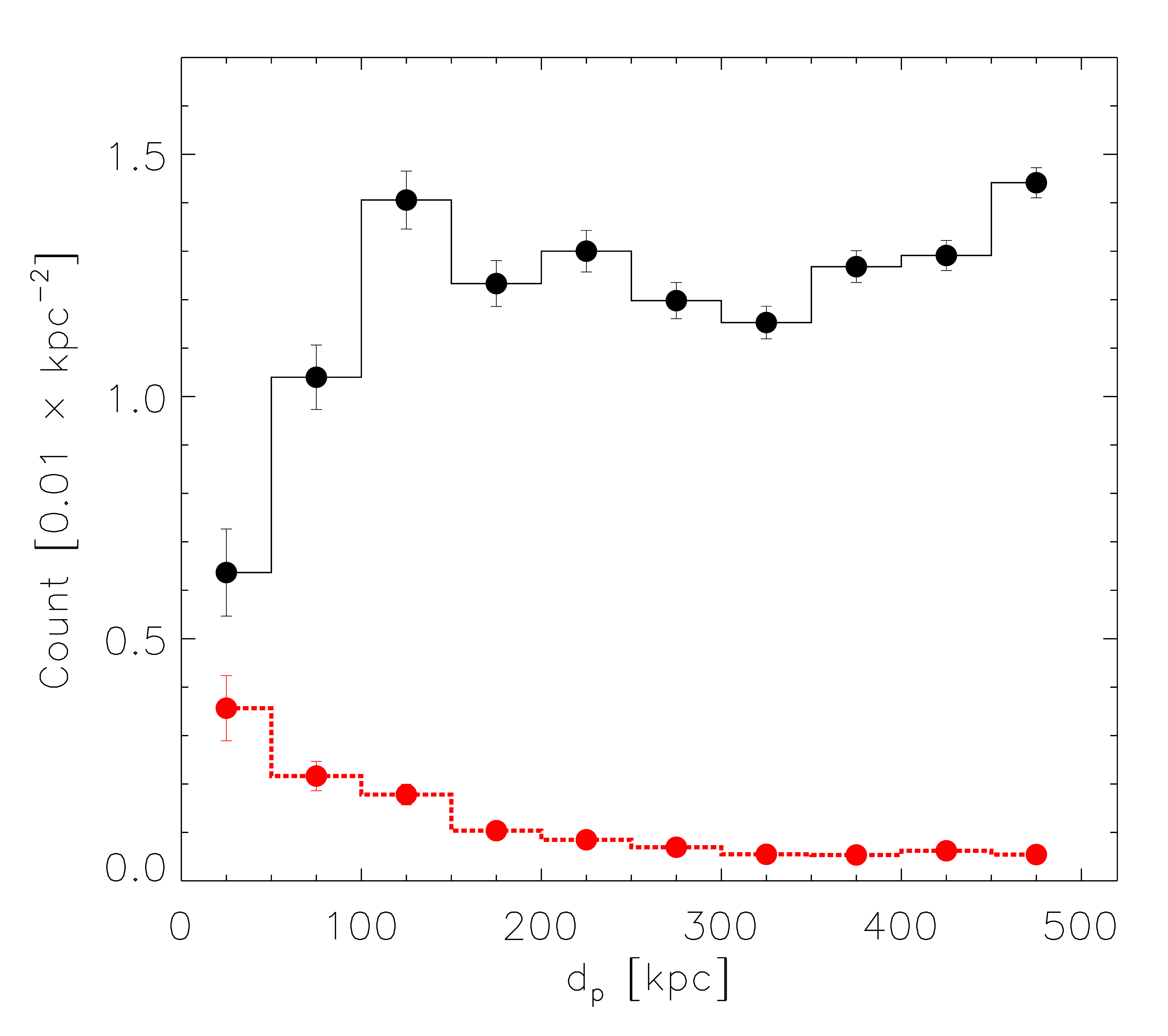}
  \caption{Distributions of projected distances of quasar pairs, which have been normalized by the circular area spanned by each annulus. Black symbols are for all the quasar pairs, and red symbols are only for quasar pairs with detected TRA \Mgii\ absorption lines in the background quasar spectra.}
  \label{fig:dp}
\end{figure}

The blueshifted and/or asymmetrical profiles of emission lines could bias quasar redshifts of the SDSS pipeline, which are able to result in an uncertainty from 100 \kms\ to 3000 \kms\ \citep[e.g.,][]{2010MNRAS.405.2302H,2011ApJS..194...45S,2019ApJS..244...36C}. The large uncertainty could give rise to an absorber detected at the quasar redshift absolutely outside galaxy halo. Therefore, the more accurate quasar redshifts are required to probe the environmental gas surrounding quasars via absorption lines with \zabs\ $\approx$ \zem. Narrow emission line often provides quasar redshift with higher accuracy than broad emission line does. And also, \citealt{2016ApJ...817...55S} suggests an uncertainty of about 200 \kms\ when quasar redshifts are determined from \Mgiiwave\ emission lines relative to those measured from \Oiiib\ emission lines. This small uncertainty suggests that the \Mgii\ emission line can better determine quasar redshift than the \lya, \Ciii, and \Civ\ emission lines do \citep[e.g.][]{2010MNRAS.405.2302H,2011ApJS..194...45S}. Therefore, we determine quasar redshifts in the following order.
\begin{enumerate}
  \item We adopt the redshifts of \cite{2010MNRAS.405.2302H} \footnote{http://das.sdss.org/va/Hewett$\_$Wild$\_$dr7qso$\_$newz/} for quasars included in the SDSS DR7Q, who have improved the redshifts for the SDSS-I/II quasars using a cross-correlation method.
  \item When narrow \oii\ or \oiii\ emission lines are available by the SDSS spectra, we determine quasar redshifts by fitting these emission lines with Gaussian function(s).
  \item Using the template of \cite{2010MNRAS.405.2302H}, we measure the redshifts from \Mgii\ or \Ciii\ emission lines when available.
\end{enumerate}
We include a velocity correction of 200 \kms\cite[][]{2016ApJ...817...55S} for the systemic redshifts determined from the \Mgii\ emission line, so that the redshifts are more consistent with those determined from the \oiii\ emission line \cite[][]{2010MNRAS.405.2302H}. Figure \ref{fig:zem} shows the redshift distributions of background quasars with black solid-line and foreground quasars with red dash-line. The red dash-line exhibits a dip at $z_{\rm em}\sim1.1$, which is caused by the decreasing sensitivity during the split of the red and blue spectrographs\footnote{ http://www.sdss.org/dr7/instruments/spectrographs}. The projected distances of quasar pairs are displayed in Figure \ref{fig:dp}.

\section{Absorption line measurements}
\label{sect:absmeasurement}
\begin{figure*}[!ht]
  \centering
  \includegraphics[width=0.93\textwidth]{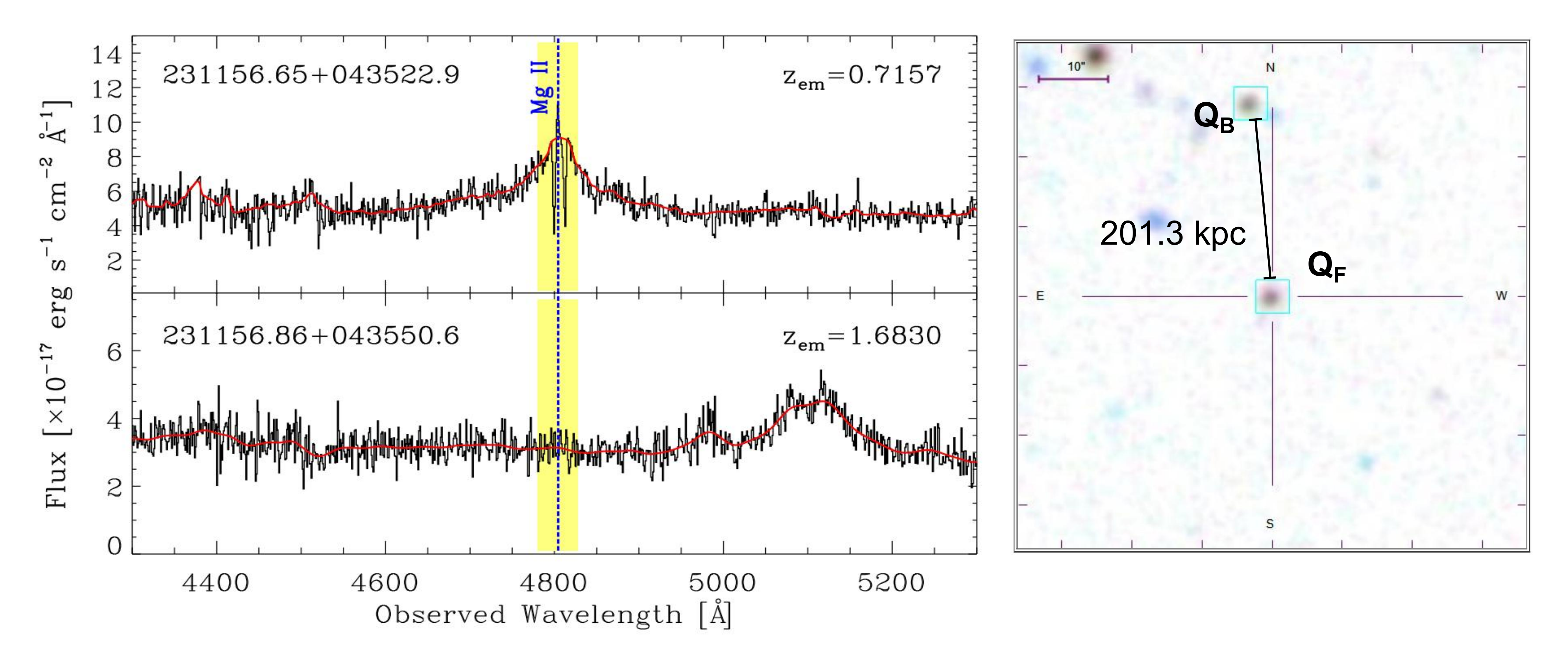}
  \includegraphics[width=0.93\textwidth]{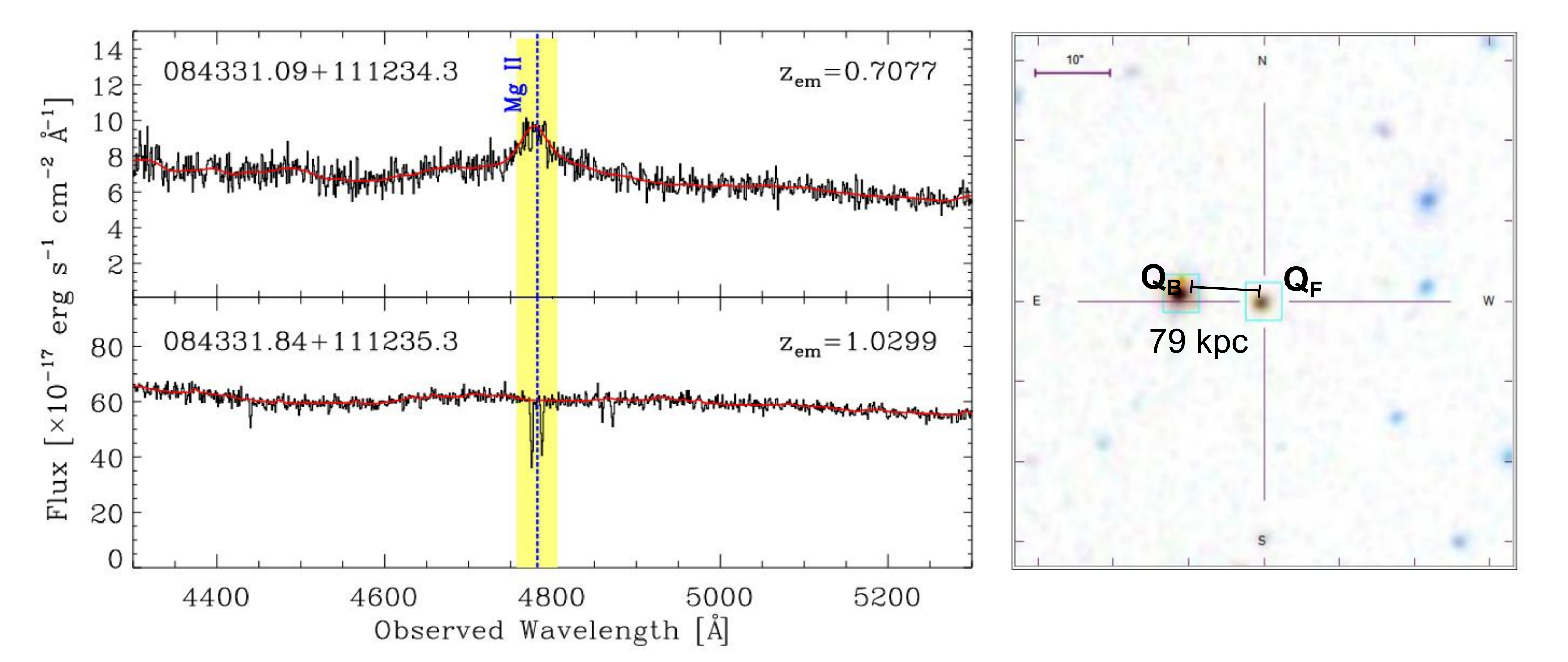}
  \caption{Examples of two pairs of quasars. Left panels show the SDSS spectra of quasars. Red solid-lines are the fitting pseudo-continuum, blue dash-lines label the positions of $\rm Mg~II$ emission lines of foreground quasars, and the shaded yellow regions label the spectral regions used to search for associated \Mgii\ absorption lines of foreground quasars. For the quasar pair of $\rm J231156.65+043522.9$ and $\rm J231156.86+043550.6$, having a projected distance of $d_{\rm p}=201.3$ \kpc, an obvious \Mgii\ absorption line is displayed in the spectra of foreground quasar, but not in the spectra of background quasar. For the quasar pair of $\rm J084331.09+111234.3$ and $\rm J084331.84+111235.3$, having a projected distance of $d_{\rm p}=79$ \kpc, an obvious \Mgii\ absorption line is displayed in the spectra of background quasar, but not in the spectra of foreground quasar. Right panels show the SDSS images of the fields around the foreground quasars. The background and foreground quasars are labelled as $\rm Q_B$ and $\rm Q_F$, respectively. The purple lines at the top left of images are 10 arcsec in length.}
  \label{fig:samples}
\end{figure*}

The \Mgii\ absorption lines imprinted in the quasar spectra can be originated in intervening galaxies that are far beyond the gravitational bound of quasars, and the gas clouds within quasar's outflows, host galaxies, halos, and galaxy clusters. Using large samples of LOS absorption lines, previous works\citep[e.g.,][]{2017ApJ...848...79C,2018ApJS..235...11C} claims that: (1) the relative velocity of $\rm \upsilon_r<6000$ \kms\ with respect to quasar systemic redshifts is a safe boundary to constrain a vast majority of \Mgii\ absorption lines from quasar's outflows, host galaxies, halos, and galaxy clusters; while (2) the velocity range of $\rm \pm 1500$ \kms\ can well constrain the \Mgii\ absorption lines formed within quasar's host galaxies, halos, and galaxy clusters. We also note from \cite{2018ApJS..235...11C} that the fraction of the intervening \Mgii\ absorption lines is similar to that of the associated ones in the velocity range of $\rm \upsilon_r = 1500$ --- 6000 \kms. In other words, the associated \Mgii\ absorption lines are significantly contaminated by the intervening ones in the velocity range of $\rm \upsilon_r = 1500$ --- 6000 \kms. In addition, the associated \Mgii\ absorption lines with $\rm \upsilon_r = 1500$ --- 6000 \kms\ are dominated by quasar's outflows. This work aims to investigate the properties of gas clouds within quasar's host galaxies and halos, making use of the \Mgii\ absorption lines with \zabs\ $\approx$ \zemfg. Therefore, in both the foreground and background quasar spectra, we search for \Mgii\ absorption lines within the spectral regions of $\rm \pm 1500$ \kms\ around \Mgii\ emission lines of the foreground quasars.  Note that, if the spectral regions used to search for \Mgii\ absorption lines are located within a broad absorption feature, we directly exclude the spectra from our parent sample. Figure \ref{fig:samples} shows the SDSS spectra and images of two pairs of quasars. The shaded yellow regions label the wavelength ranges used to search for associated \Mgii\ absorption lines of foreground quasars.

We firstly masked out the strong absorption lines around the spectral regions used to searched for \Mgii\ absorption lines. Then cubic spline (for underlying continuum) plus Gaussian functions (for emission line features) are invoked to derive the pseudo-continuum fitting (see the red-solid lines shown in Figure \ref{fig:samples}), which is used to normalize spectral flux (flux divided by the pseudo-continuum fitting) and flux uncertainty (flux uncertainty divided by the pseudo-continuum fitting) \citep[e.g.,][]{2015ApJS..221...32C,2018ApJS..235...11C}. Here the spline is extrapolated over the continuum underneath the emission line.
We search for and visually inspect the \Mgii\ absorption doublets on the normalized spectra, and model each doublet with a pair of Gaussian functions. The integration of the Gaussian modeling gives absorption equivalent width at rest frame (\Wr). The error of the \Wr\ ($\sigma_{\rm w}$) is derived by directly integrating the normalized flux uncertainty within $\pm3\sigma$ widths, where the $\sigma$ is given by the Gaussian modeling. We only keep the \Mgii\ absorption lines with $W_{\rm r}^{\rm \lambda2796}\ge2\sigma_{\rm w}$ and $W_{\rm r}^{\rm \lambda2796}\ge0.15$ \AA. For the spectra without detected \Mgii\ absorption lines, we estimate the $\sigma_{\rm w}$ within 200 \kms, which are taken as the upper limits of the \Wr\ as well. The absorption parameters of the detected \Mgii\ absorption lines are listed in \ref{tab:fgabs} and \ref{tab:bgabs}.

\section{Transverse absorption lines}\label{sect:TRA}
\begin{figure}[!ht]
  \centering
  \includegraphics[width=0.47\textwidth]{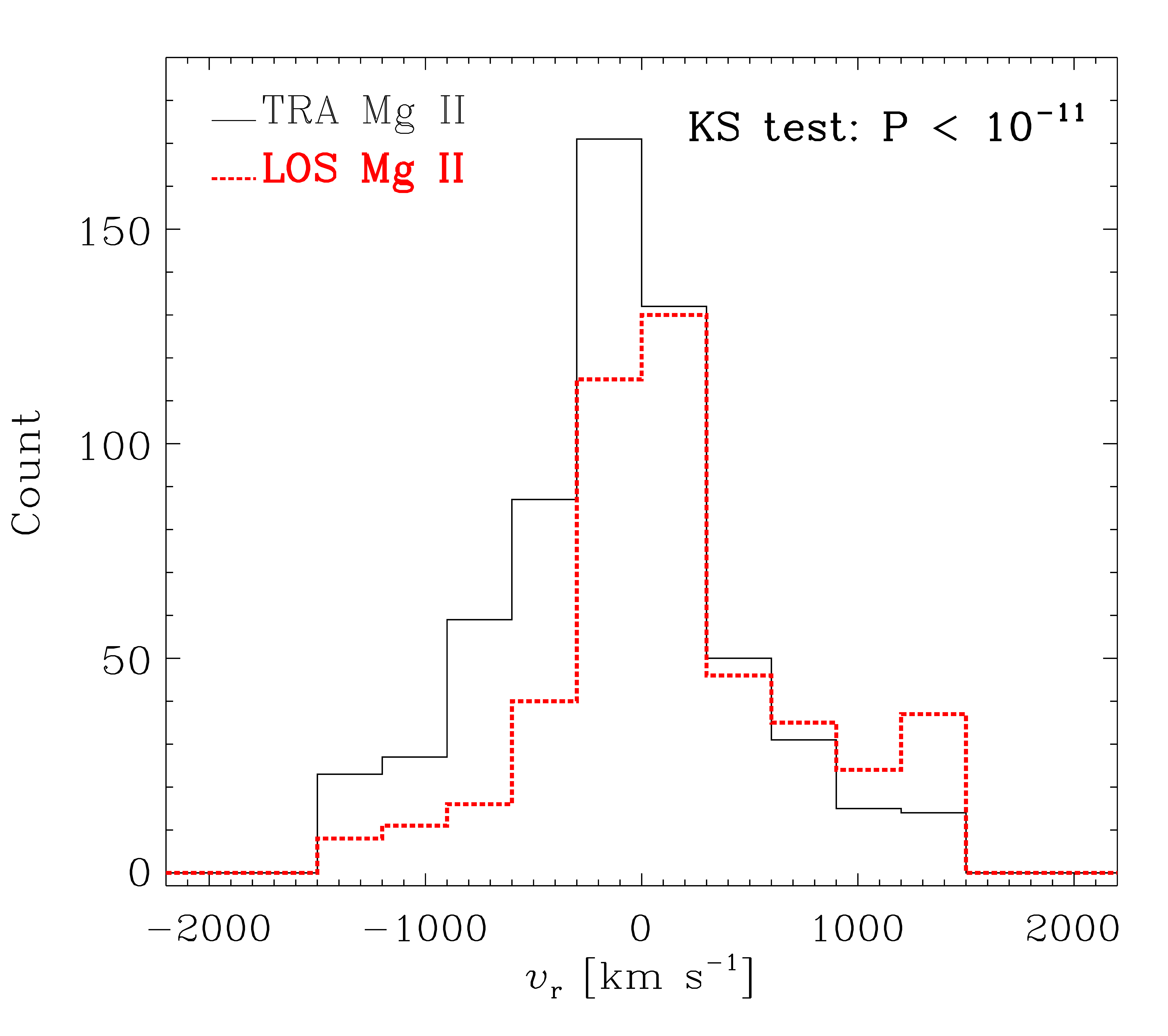}
  \caption{Velocity distributions of \Mgii\ absorption lines relative to foreground quasar systemic redshifts. Black solid-line and red dash-line are for the TRA and LOS \Mgii\ absorption lines, respectively.}
  \label{fig:vr}
\end{figure}

In the large sample of 10025 foreground-background quasar pairs, we find that there are 598 background quasars with detected TRA \Mgii\ absorption lines that are associated to foreground quasars. Figure \ref{fig:vr} shows the velocities of the TRA \Mgii\ absorption lines relative to foreground quasar systemic redshifts \footnote{$\upsilon_r = \frac{(1+z_{\rm abs})^2-(1+z_{\rm em})^2}{(1+z_{\rm abs})^2+(1+z_{\rm em})^2}$, where \zabs\ is the absorption line redshift determined from \Mgiia, and \zem\ is the systemic redshift of foreground quasar.}. About 90\% of the TRA \Mgii\ absorption lines are located within $|\upsilon_r|<600$ \kms\ ($0.002c$). The projected distances of the TRA \Mgii\ absorption lines are also displayed with red dash-line in Figure \ref{fig:dp}. It is clearly seen that the TRA \Mgii\ absorption lines are more likely to be detected in the spectra of background quasars with smaller projected distances. In addition, the projected distances of the TRA \Mgii\ absorbing clouds are from 20 \kpc\ to 500 \kpc, which implies that the TRA \Mgii\ absorption lines would be mainly dominated by the halos of foreground quasar host. Of course, neighbouring halos of quasars might also contribute to the TRA \Mgii\ absorption lines with $d_{\rm p}>300$ \kpc\ \citep[e.g.,][]{2014MNRAS.439.3139Z}.

\begin{figure}[!ht]
  \centering
  \includegraphics[width=0.47\textwidth]{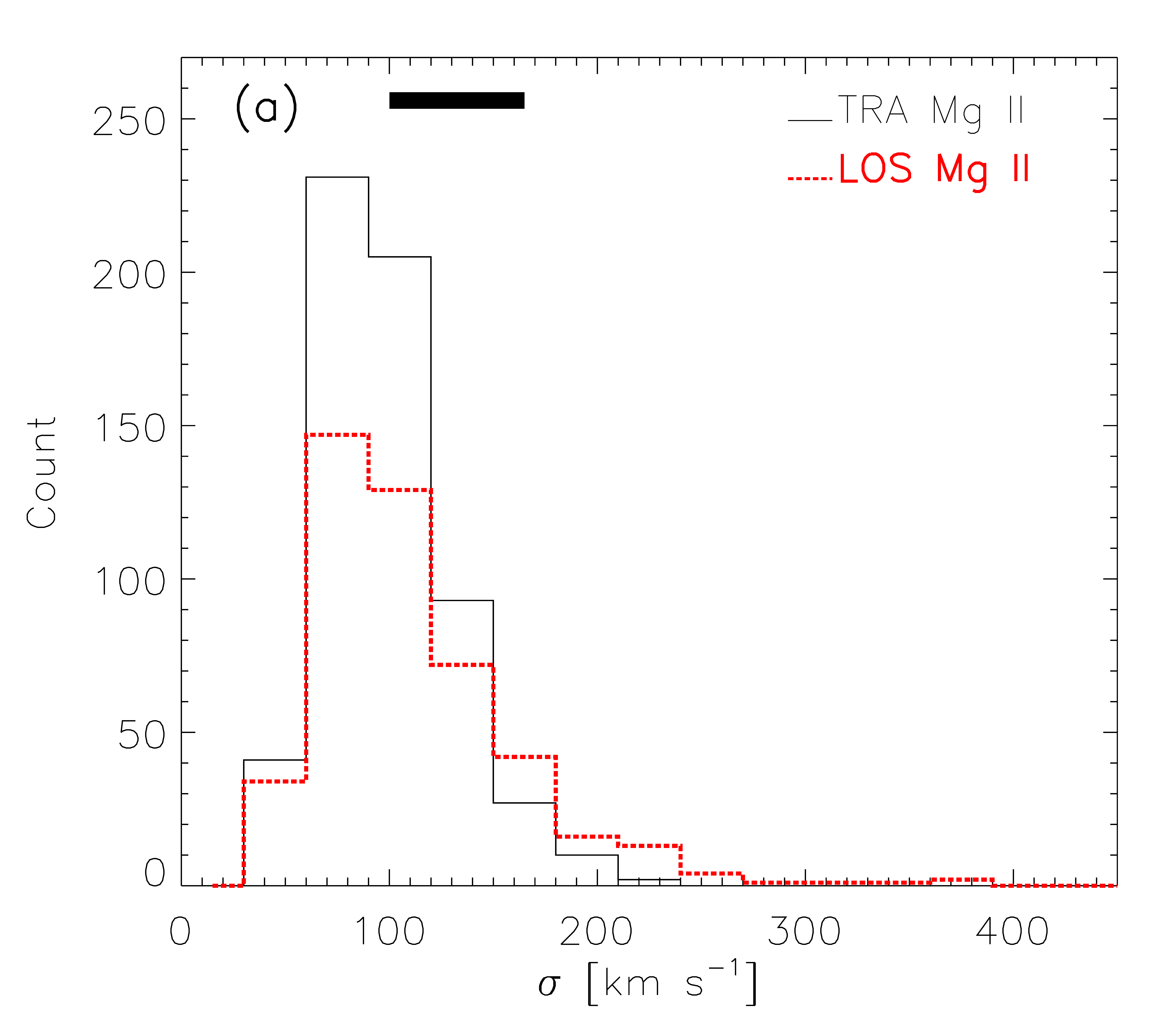}
  \includegraphics[width=0.47\textwidth]{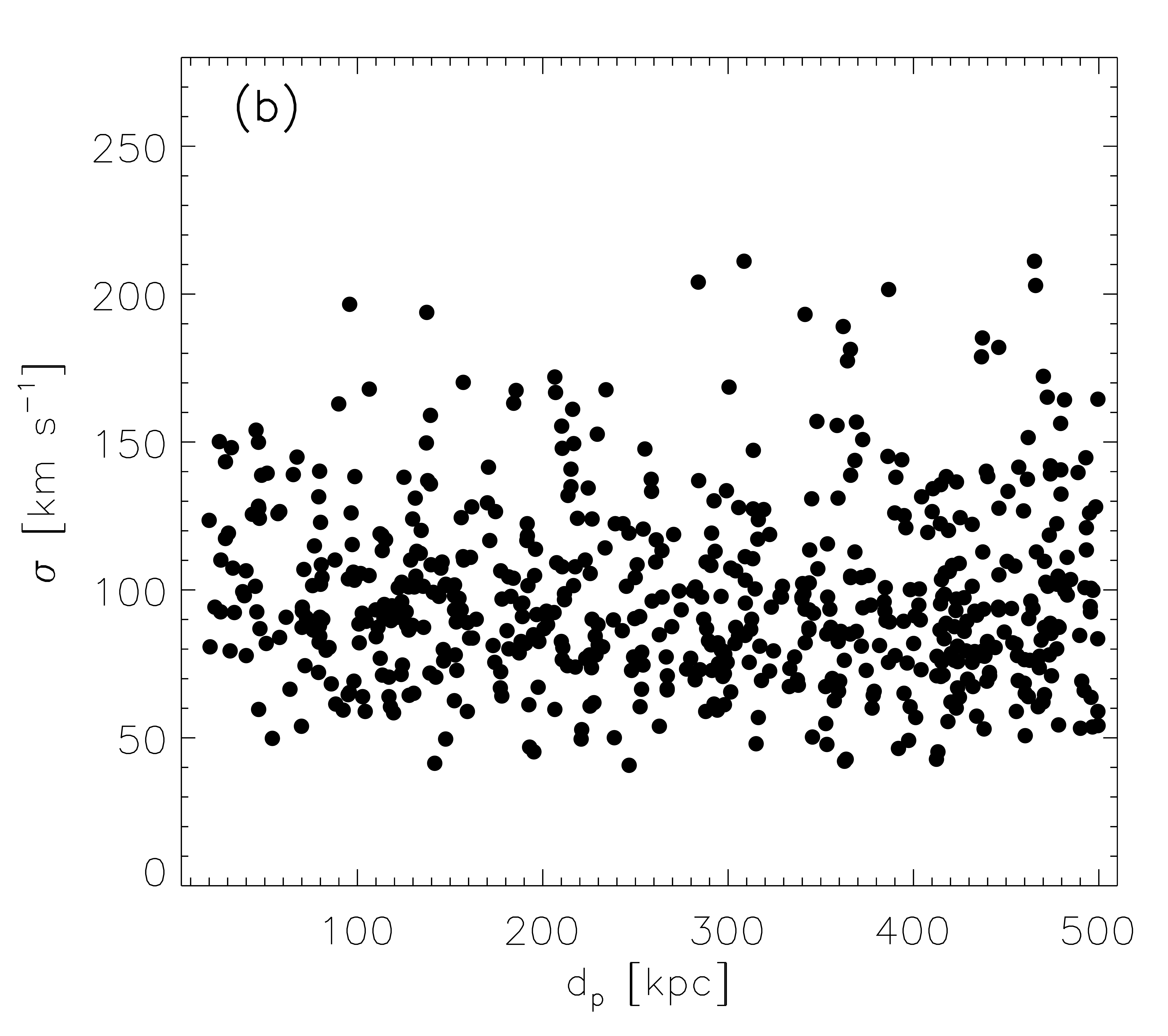}
  \caption{Panel (a): Velocity dispersions of gas clouds traced by \Mgii\ absorption lines. Black solid-line and red dash-line are for the TRA and LOS \Mgii\ absorption lines, respectively. The thick horizontal line is for the mean uncertainty of line widths, accounting for the typical spectral resolution of the SDSS/BOSS. Panel (b): Velocity dispersions of gas clouds traced by TRA \Mgii\ absorption lines as a function of projected distances.}
  \label{fig:fwhm}
\end{figure}

The velocity widths of \Mgii\ absorption lines provide dynamical information of absorbing clouds around quasar host galaxies. For the luminous red galaxies (LRGs), \cite{2014MNRAS.439.3139Z} revealed that the velocity widths of \Mgii\ absorption lines with $\sigma<300$ \kms\ on scales $d_p<300$ \kpc, where the $\sigma$ is the velocity dispersion of gas cloud measured from the Gaussian fits of absorption lines, mainly reflect the motion of gas clouds within halos of LRGs host, while the velocity widths of \Mgii\ absorption lines with $\sigma>300$ \kms\ on scales $d_p>600$ \kpc\ are related to the motion of gas clouds of the neighbouring halos of the LRGs. In Figure \ref{fig:fwhm}, we show the velocity dispersions measured from the TRA \Mgii\ absorption lines. It can be seen that the velocity dispersions of the \Mgii\ gas clouds are less than 250 \kms, and mainly constrained within a velocity range of 60 --- 150 \kms\ (about 87\% of the TRA \Mgii\ absorption lines). Meanwhile, velocity dispersions are not obviously changed with projected distances. The velocity dispersion with small/mild values and not changed with projected distances, imply again that the TRA \Mgii\ absorption lines would be mainly reflect the properties of gas clouds within the halos of quasar host, but not within the neighbouring halos of quasars.

\begin{figure*}[!ht]
  \centering
  \includegraphics[width=0.9\textwidth]{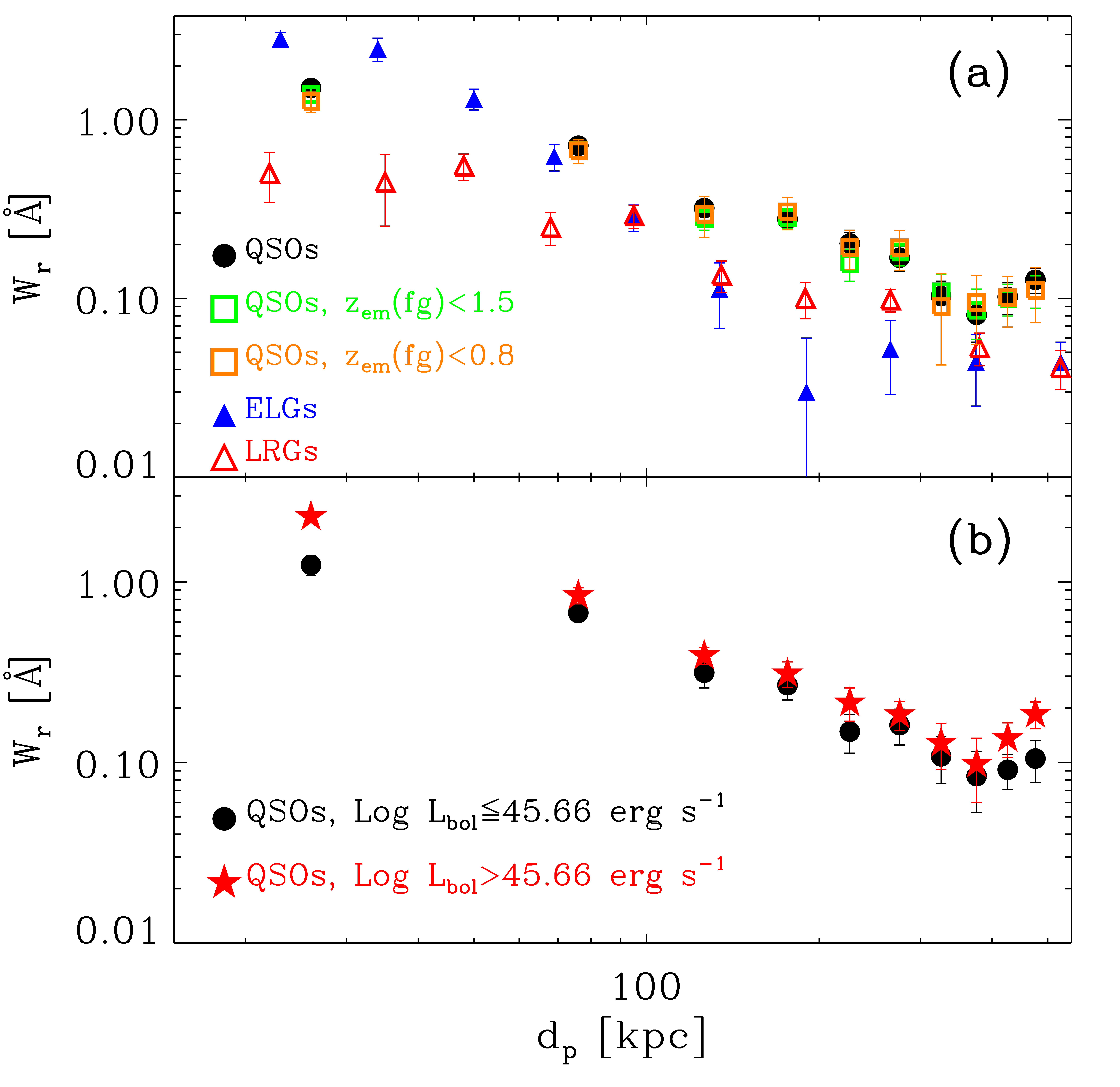}
  \caption{\Mgii\ rest equivalent width, the sum of \Mgiia\ and \Mgiib\ lines, as a function of projected distance for the TRA absorption lines. Panel (a): Black filled circles are for all the quasar pairs; green and orange unfilled squares are for the quasar pairs whose foreground quasars have $z_{em}(fg)<1.5$ and $z_{em}(fg)<0.8$, respectively; blue filled triangles and red unfilled triangles are for the TRA \Mgii\ absorption lines around the emission line galaxies and luminous red galaxies from \cite{2018ApJ...866...36L}, respectively. Panel (b): Black circles are for the quasar pairs whose foreground quasars have $Log L_{\rm bol}\le45.66$ \ergs, and the red stars are for the quasar pairs whose foreground quasars have $Log L_{\rm bol}>45.66$ \ergs, where the $Log L_{\rm bol}=45.66$ \ergs\ is the median value of the bolometric luminosities of foreground quasars.}
  \label{fig:dist_w}
\end{figure*}

For the background quasars, we shift them to rest frame of the foreground quasars or the \Mgii\ absorbers when the TRA \Mgii\ absorption lines are available, and then produce median composite spectra. We measure the equivalent widths of \Mgii\ absorption lines in composite spectra. The results are shown in Figure \ref{fig:dist_w} and listed in Table \ref{tab:wr}. The \Mgii\ equivalent width is decreased with increasing projected distance before $d_{\rm}<400$ \kpc. We invoke the Spearman's correlation test to the relationship between the \Mgii\ equivalent widths and projected distances, which yields a correlation coefficient $\rho=-0.915$ and a probability $P < 10^{-4}$ of no correlation. This decreased trend is similar to previous results and also hosted by other transitions, such as \Civ, \Cii, \Siiv\ absorption lines, no matter what type of galaxies these absorbing clouds surround \citep[e.g.,][]{2010ApJ...717..289S,2014ApJ...796..140P,2014MNRAS.439.3139Z,2015MNRAS.452.2553J,2016MNRAS.457..267L,2018ApJ...866...36L}.

Recently, using background quasars, \cite{2018ApJ...866...36L} investigated the properties of \Mgii\ absorption lines around emission line galaxies (ELGs) and luminous red galaxies (LRGs). In Figure \ref{fig:dist_w}, we also display the results of \cite{2018ApJ...866...36L} with blue or red triangles. The \Mgii\ absorption strengths are stronger around ELGs than around quasars on scales smaller than 50 \kpc. This might be due to some of the \Mgii\ absorption lines around ELGs associated to the outflows of ELGs on small scales, while the TRA \Mgii\ absorption lines of quasars are not related to outflows, since the the vast majority of quasar sightlines would be close to the axis of accretion disk. We also note that the \Mgii\ absorption strengths are stronger around quasars than around LRGs on all scales, and than around ELGs on scales greater than 100 \kpc. The ELGs and LRGs used by \cite{2018ApJ...866...36L} have redshift ranges of $z\approx0.4$ --- 1.5, and $z\approx0.4$ --- 0.8, respectively. These redshift ranges are obviously smaller than that ($z\approx0.4$ --- 2.6) of foreground quasars used in this paper. Here we limit our quasar pair sample with redshifts of foreground quasars $z_{\rm em}(fg)<1.5$ and $z_{\rm em}(fg)<0.8$, respectively. The results are displayed with green and orange squares in Figure \ref{fig:dist_w}, respectively. It is clearly seen that the discrepancies between quasars and ELGs or LRGs are not caused by redshifts. In order to check the influence from quasar radiations, we divide our quasar pair sample into two subsamples, namely, the subsamples of quasars with lower and higher luminosities, respectively. The results are also shown in Figure \ref{fig:dist_w}, which suggests that the \Mgii\ equivalent width of the quasars with higher luminosities would be slightly greater than that of the quasars with lower luminosities.

\begin{table}
\caption{The \Mgii\ equivalent widths measured from composite spectra} \tabcolsep 3mm \centering \label{tab:wr}
 \begin{tabular}{ccccccc}
 \hline\hline\noalign{\smallskip}
$d_p$ bin& $N_{\rm pair}$ & $W_{\rm r}^{\lambda2796}$ & $W_{\rm r}^{\lambda2803}$  \\
\kpc &  & \AA & \AA \\
\hline\noalign{\smallskip}
(0	,50]	&	51	&	0.855$\pm$0.084 	&	0.647$\pm$0.059 	\\
(50	,100]	&	244	&	0.417$\pm$0.044 	&	0.298$\pm$0.037 	\\
(100,150]	&	552	&	0.185$\pm$0.027 	&	0.135$\pm$0.019 	\\
(150,200]	&	678	&	0.147$\pm$0.027 	&	0.131$\pm$0.022 	\\
(200,250]	&	915	&	0.106$\pm$0.023 	&	0.097$\pm$0.019 	\\
(250,300]	&	1033&	0.097$\pm$0.021 	&	0.072$\pm$0.017 	\\
(300,350]	&	1176&	0.056$\pm$0.015 	&	0.047$\pm$0.016 	\\
(350,400]	&	1491&	0.049$\pm$0.017 	&	0.032$\pm$0.017 	\\
(400,450]	&	1722&	0.050$\pm$0.015 	&	0.052$\pm$0.014 	\\
(450,500]	&	2149&	0.068$\pm$0.016 	&	0.059$\pm$0.013 	\\
\noalign{\smallskip}
\hline\hline\noalign{\smallskip}
\end{tabular}
\end{table}

\section{Line-of-sight absorption lines}\label{sect:LOS}
Absorption features detected in foreground quasar spectra with \zabs\ $\approx$ \zem\ are mainly formed within quasar's outflows/winds, host galaxies, halos, and galaxy clusters. Broad absorption lines (BALs) with line width larger than 2000 \kms, and mini-BALs with line width larger than a few hundreds \kms\ and less than 2000 \kms\ are usually considered to be associated to quasar's outflows/winds. Meanwhile, we note that for LRGs, the line width is less than 400 \kms\ for the \Mgii\ absorption lines formed within the gas clouds with projected distances less 1 \mpc\ \citep[e.g.,][]{2014MNRAS.439.3139Z}. In this work, we limit the TRA \Mgii\ absorption lines with $d_{\rm p}<500$ \kpc\ and mainly focus on the absorption lines formed within the gas clouds of quasar's host galaxies and halos. Therefore, we only retain the LOS \Mgii\ absorption lines with line width less than 500 \kms. In the large sample of 10025 foreground-background quasar pairs, we find that there are 454 foreground quasars with detected LOS \Mgii\ absorption lines.

The velocity dispersions measured from the LOS \Mgii\ absorption lines are shown with red dash-line in Figure \ref{fig:fwhm}. In addition to 4 LOS \Mgii\ absorption lines, all the LOS \Mgii\ absorption lines yield a velocity dispersion $\sigma<300$ \kms. About 75\% of the LOS \Mgii\ absorption lines are also constrained within a velocity range of 60 --- 150 \kms, which implies that most of the LOS \Mgii\ gas clouds could be gravitationally bounded by quasar halos.

The velocities of LOS \Mgii\ absorption lines relative to foreground quasar systemic redshifts are shown with red dash-line in Figure \ref{fig:vr}. About 79\% of the LOS \Mgii\ absorption lines are located within $|\upsilon_r|<600$ \kms. We perform the Kolmogorov-Smirnov (KS) test to the velocity distributions of the TRA and LOS \Mgii, which yields a probability of $P<10^{-11}$. This small probability suggests that the velocities of the TRA \Mgii\ are obviously different from those of the LOS ones. As a whole, the TRA and LOS \Mgii\ absorption lines have mean velocities of $\upsilon_r=-116$ \kms\ and 160 \kms, when compared to quasar systemic redshifts. This implies that the TRA \Mgii\ are slightly redshifted and the LOS ones are slightly blueredshifted relative to quasar systems. \cite{2018ApJ...857..126L} also claimed that the TRA absorption lines exhibit a small redshift relative to quasar systems. Both the redshifted TRA and blueshifted LOS absorption lines imply that the absorbing gas is on average dominated by galaxy's outflows. While, the small values of redshift (-116 \kms) and blueshift (160 \kms) can be compared to the uncertainty of quasar systemic redshits. Therefore, here we can not further investigate the intrinsic origins of the TRA and LOS \Mgii\ with present data.

\section{Incident rate of \Mgii\ absorption lines}\label{sect:composite}
We define the incident rate of \Mgii\ absorption lines at different projected distance as follows:
\begin{equation}
\label{eq:fc}
f_c = \frac{N_{\rm abs}}{N_{\rm QSO}},
\end{equation}
where, at a given range of projected distance, $N_{\rm abs}$ represents the number of detected \Mgii\ absorption lines (See Section \ref{sect:absmeasurement} for the limits of detected absorbers), and $N_{\rm QSO}$ represents the number of quasars used to search for corresponding absorption lines. The error of $f_c$ can be derived from the binomial statistics. In this section, we statistically analyze the incident rates of LOS and TRA \Mgii\ absorption lines, respectively.

For transverse absorptions, we calculate $f_c$ with Equation (\ref{eq:fc}) in several bins of projected distance, here $N_{\rm abs}$ is the number of detected TRA absorbers, and $N_{\rm QSO}$ is the number of background quasars used to search for TRA \Mgii\ absorption lines. The results are listed in Table \ref{tab:fc}.

The distance between the LOS absorbing cloud and quasar central region is a mystery, so we have no knowledge of the relationship between the $f_c$ of the LOS \Mgii\ absorption line and distance. While, we still estimate the $f_c$ of LOS \Mgii\ absorption lines with Equation (\ref{eq:fc}), where $N_{\rm abs}$ represents the number of detected LOS \Mgii\ absorption lines, and $N_{\rm QSO}$ represents the number of foreground quasar used to search for LOS \Mgii\ absorption lines. In this way, the derived $f_c$ is not the function of distance, and should be a constant.

\begin{figure*}[!ht]
  \centering
  \includegraphics[width=0.9\textwidth]{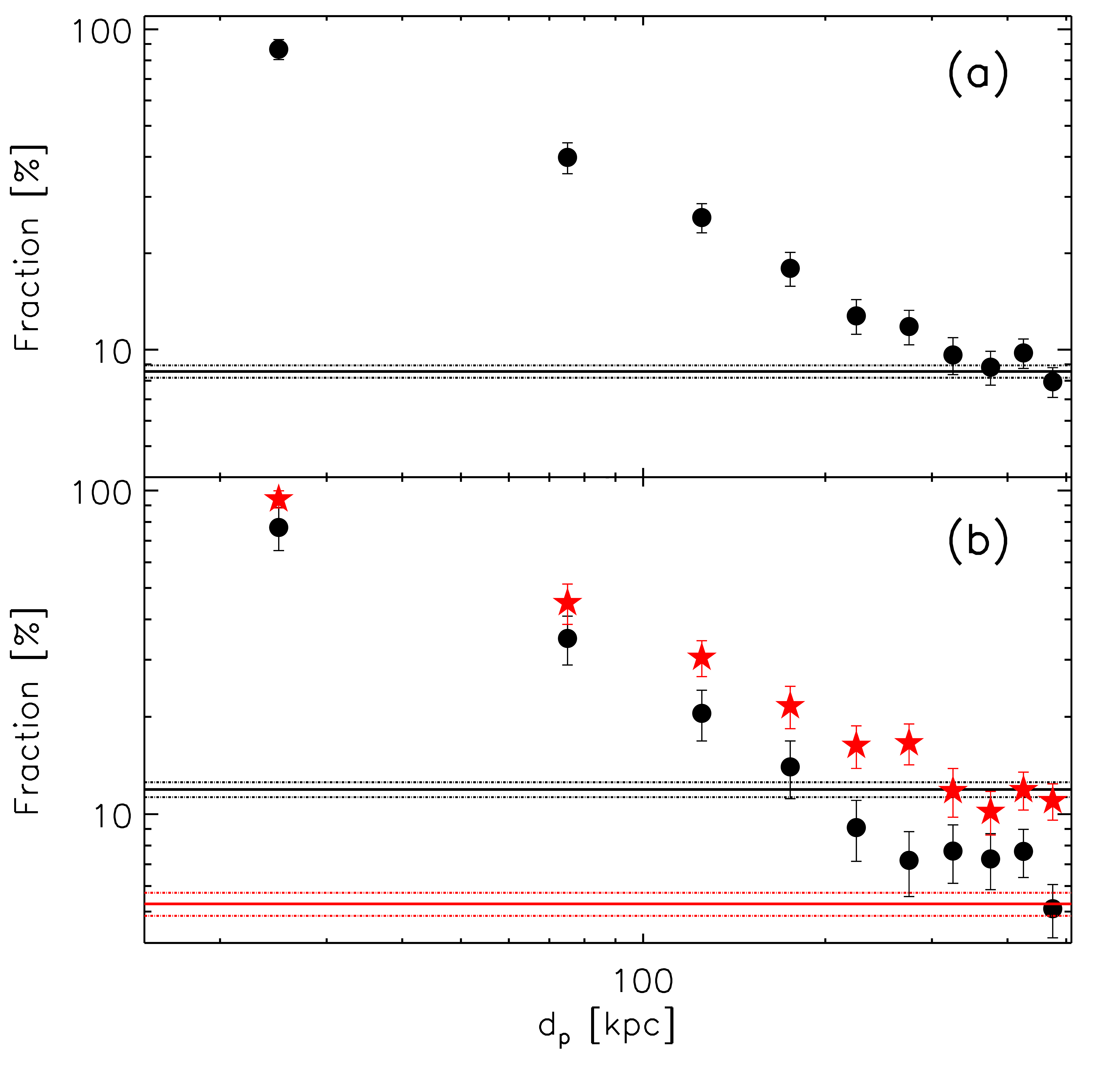}
  \caption{The incident rate of detected TRA \Mgii\ absorption lines as a function of projected distance. Horizontal solid lines represent incident rates of detected LOS \Mgii\ absorption lines, and dot-dash lines represent  $\pm1 \sigma$ error of the incident rates of detected LOS \Mgii\ absorption lines. Note that these fractions for the LOS \Mgii\ are averaged over the entire velocity bin considered. Panel (a) is for all the quasar pairs, where $f_{\rm c} = 8.554\pm0.381\%$ for the LOS \Mgii\ absorption lines. Panel (b): Black symbols (circles and horizonal lines) are for the quasar pairs whose foreground quasars have $Log L_{\rm bol}\le45.66$ \ergs, and the red symbols (stars and horizonal lines) are for the quasar pairs whose foreground quasars have $Log L_{\rm bol}>45.66$ \ergs. For the LOS \Mgii\ absorption lines, faint foreground quasars have $f_{\rm c} = 11.919\pm0.633\%$, and luminous quasars have $f_{\rm c} = 5.283\pm0.433\%$.}
  \label{fig:fc}
\end{figure*}

In panel (a) of Figure \ref{fig:fc}, we show the derived $f_c$ as a function of projected distance. It is clearly seen that the incident rate of the TRA absorption lines is decreased quickly with increasing projected distance. We invoke the Spearman's correlation test to the relationship between the $f_c$ and projected distances, which yields a correlation coefficient $\rho=-0.964$ and a probability $P < 10^{-6}$ of no correlation. The decreasing tendency is consistent with previous results \citep[e.g.,][]{2014MNRAS.441..886F,2014ApJ...796..140P,2015MNRAS.452.2553J}. It is interesting that the $f_c$ of TRA \Mgii\ absorption lines is significantly ($>4\sigma$) higher than that of LOS absorption lines on scales smaller than 200 \kpc, while the TRA and LOS \Mgii\ have similar (within $3\sigma$) incidences on larger scales. This could be an indication that the quasar radiation is not significantly impacting the gas halos of quasars at scales $d_p>200$ \kpc. On small scales, significant difference in the incident rates between TRA and LOS \Mgii\ absorption lines is likely related to the anisotropic radiation originated from quasars \citep[e.g.,][]{2013ApJ...776..136P,2016ApJS..226...25L,2018ApJ...857..126L,2019ApJ...884..151J}. Quasar's outflows/jets might inject significant energy into surrounding environment, through kinetic feedback and/or radiative pressure. Therefore, the powerful outflows/jets of quasars would over ionize the medium with lower ionization state, which is also the most possible interpretation for the proximity effect along the line of sights. While, the illumination from quasars along the transverse directions would much less than that along the sightlines of quasars \citep[e.g.,][]{2019ApJ...884..151J}, so there is an inverse proximity effect along the transverse directions (commonly called as transverse proximity effect).

\begin{table}
\caption{The incident rate of TRA \Mgii\ absorption lines} \tabcolsep 11mm \centering \label{tab:fc}
 \begin{tabular}{ccccccc}
 \hline\hline\noalign{\smallskip}
$d_p$ bin (\kpc)& $f_{\rm c}$  \\
\hline\noalign{\smallskip}
(0	,50]	&	86.7$\pm$6.2 	\\
(50	,100]	&	39.8$\pm$4.4 	\\
(100,150]	&	25.9$\pm$2.7 	\\
(150,200]	&	17.9$\pm$2.2 	\\
(200,250]	&	12.8$\pm$1.6 	\\
(250,300]	&	11.8$\pm$1.5	\\
(300,350]	&	9.6 $\pm$1.3 	\\
(350,400]	&	8.8 $\pm$1.1 	\\
(400,450]	&	9.8 $\pm$1.0	    \\
(450,500]	&	7.9 $\pm$0.8 	\\
\noalign{\smallskip}
\hline\hline\noalign{\smallskip}
\end{tabular}
\end{table}

In order to check the influence on the incident rate of \Mgii\ absorption lines from quasar radiation, we compare the incident rates of \Mgii\ absorption lines between the subsamples whose foreground quasars host different luminosities. The results are displayed in Figure \ref{fig:fc} as well. It is clearly seen that the quasars with higher luminosities (red lines in panel (b) of Figure \ref{fig:fc}) have an obviously ($>5\sigma$) smaller incidence of LOS \Mgii\ absorption lines, when compared to the quasars with lower luminosities (black lines in panel (b) of Figure \ref{fig:fc}). This difference would be the expectation of the proximity effect along quasar sightline, since the quasars with higher luminosities evaporate more \Mgii\ absorbing gas close to quasars, when compared to the quasars with lower luminosities. While along the transverse direction, it can be clearly seen from Figure \ref{fig:fc} that the incident rate of TRA \Mgii\ absorption lines is higher for more luminous quasars (red stars in panel (b) of Figure \ref{fig:fc}). It is believe that only a small portion of quasar's illumination pass through the gas clouds along transverse direction. Therefore for the more luminous quasars, not only the \Mgii\ absorbing gas is not evaporated but also more neutral magnesium is ionized into ionic gas. This leads to a greater incident rate of TRA \Mgii\ absorption lines for the more luminous quasars. Of course, these discrepancies could be also related to a bias in the quasar host properties. For instance, the bright quasars would be expected to host halos with masses larger than those of faint quasars, which could also affect in the gas distribution.

\section{Summary} \label{sec:summary}
We have constructed a large sample of 10025 quasar pairs with $d_{\rm p}<500$ \kpc\ from the final data set of the SDSS-IV, to analyze the properties of \Mgii\ absorption lines around foreground quasars, including both the LOS and TRA absorptions. We find that there are 598 background quasars with detected TRA \Mgii\ absorption lines, and 454 foreground quasars with detected LOS \Mgii\ absorption lines.

The velocity dispersion ($\sigma$, line width) of all the TRA \Mgii\ absorption lines is less than 250 \kms. The line width of LOS \Mgii\ absorption lines have a value of $\sigma<300$ \kms, except for 4 systems with $300<\sigma<400$ \kms. Both the LOS and TRA \Mgii\ absorption lines are mainly constrained within a velocity dispersion of $60<\sigma<150$ \kms. Meanwhile, for a vast majority of both the TRA and LOS \Mgii\ absorption lines, the relative velocity of absorption lines ($\upsilon_r$, with respect to foreground quasar systemic redshift) is within a range of $|\upsilon_r|<600$ \kms. Combining the small/mild velocity dispersion and small relative velocity, most of the TRA and LOS \Mgii\ absorption lines would be dominated by the gas medium gravitationally bounded by quasar halo.

Both the equivalent width and incident rate of TRA \Mgii\ absorption lines are rapidly decreased with increasing distance. The incident rate of TRA \Mgii\ absorption lines is obviously ($>4\sigma$) greater than that of LOS \Mgii\ absorption lines at projected distances $d_p<200$ \kpc, and they become consistent within $3\sigma$ at $d_p>200$ \kpc. The anisotropic radiation from quasars would be the most possible interpretation for the anisotropic absorption around quasars. The powerful outflow/jet likely over ionizes the gas medium with low ionization state along quasar sightlines, and impose a so-called proximity effect that obviously reduces the incidence of absorptions along quasar sightlines. While along the transverse direction, the less illumination from quasars can not drive a transverse proximity effect. The TRA and LOS \Mgii\ absorption lines both have similar incidences at $d_p>200$ \kpc, which hints that the quasar feedback from powerful outflows/jets is not significantly influencing the \Mgii\ gas around quasars on large scales. Of course, the scale of quasar feedback would be related to quasar's luminosity or accretion rate.

\section*{Acknowledgements}
We deeply thank the anonymous referees for her/his helpful and careful comments. This work is supported by the National Natural Science Foundation of China (12073007), the Guangxi Natural Science Foundation (2019GXNSFFA245008; GKAD19245136), and the Scientific Research Project of Guangxi University for Nationalities (2018KJQD01).

\bibliography{cG}{}
\bibliographystyle{aasjournal}



\end{document}